\documentclass[journal]{vgtc}                
\ifpdf
  \pdfoutput=1\relax                   
  \usepackage{graphicx}                
  \DeclareGraphicsExtensions{.pdf,.png,.jpg,.jpeg} 
\else
  \usepackage{graphicx}                
  \DeclareGraphicsExtensions{.eps}     
\fi%

\graphicspath{{figures/}{pictures/}{images/}{./}} 

\usepackage{microtype}                 
\PassOptionsToPackage{warn}{textcomp}  
\usepackage{textcomp}                  
\usepackage{mathptmx}                  
\usepackage{times}                     
\usepackage{cite}                      
\usepackage{tabu}                      
\usepackage{booktabs}                  
\usepackage{color}
\usepackage{comment}
\usepackage[inline]{enumitem}
\usepackage{wrapfig}                   

\usepackage{xspace}
\usepackage{balance}

\onlineid{0}

\vgtccategory{Research}
\vgtcpapertype{application/design study}

\newcommand{\jdf}[1]{{\textcolor{cyan}{\tiny[\textbf{JDF:} #1]}}}

\newcommand{\todo}[1]{\marginpar{\textcolor{red}{\tiny #1}}}

\newcommand{\eg}{e.g.,\xspace}
\newcommand{\ie}{i.e.,\xspace}

\newcommand{\paovis}{PAOHVis\xspace}
\newcommand{\rev}[1]{\textcolor{black}{#1}}
\newenvironment{revs}{\bgroup\color{black}}{\egroup}
\newcommand{\del}[1]{}

\title{Integrating Prior Knowledge in Mixed-Initiative\texorpdfstring{\\}{ }Social Network Clustering}
\author{Alexis Pister, Paolo Buono, Jean-Daniel Fekete, \textit{Senior Member, IEEE}, Catherine Plaisant, and Paola Valdivia}
\authorfooter{
\item
 Alexis Pister, Paola Valdivia, Jean-Daniel Fekete are with Universit\'e Paris-Saclay, CNRS, Inria, LRI, France. \protect\\
 E-mail: \textit{first-name.last-name}@inria.fr
 \item  Paolo Buono is with the University of Bari, Italy.\protect\\
    E-mail: paolo.buono@di.uniba.it
\item
 Catherine Plaisant is with Universit\'e Paris-Saclay, CNRS, Inria, LRI, France and University of Maryland, USA. \protect\\
 E-mail: plaisant@umd.edu
}

\shortauthortitle{Pister \MakeLowercase{\textit{et al.}}: Prior Knowledge in Network Clustering}

\abstract{%
We propose a new \rev{approach}---called PK-clustering---to help social scientists create meaningful clusters in social networks. Many clustering algorithms exist but most social scientists find them difficult to understand, and tools do not provide any guidance to choose algorithms, or to evaluate results taking into account the \emph{prior knowledge} of the scientists. Our work introduces a new clustering \rev{approach} and a visual analytics user interface that address this issue. It is based on a process that 1) captures the prior knowledge of the scientists as a set of incomplete clusters, 2) runs multiple clustering algorithms (similarly to \emph{clustering ensemble} methods), 3) visualizes the results of all the algorithms ranked and summarized by how well each algorithm matches the prior knowledge, 4) evaluates the consensus between user-selected algorithms and 5) allows users to review details and iteratively update the acquired knowledge. We describe our \rev{approach} using an initial functional prototype, then provide two examples of use and early feedback from social scientists. We believe our clustering \rev{approach} offers a novel constructive method to iteratively build knowledge while avoiding being overly influenced by the results of often randomly selected black-box clustering algorithms.
} 


\CCScatlist{ 
 \CCScat{K.6.1}{Management of Computing and Information Systems}%
{Project and People Management}{Life Cycle};
 \CCScat{K.7.m}{The Computing Profession}{Miscellaneous}{Ethics}
}

\teaser{
  \centering
  \includegraphics[width=\linewidth]{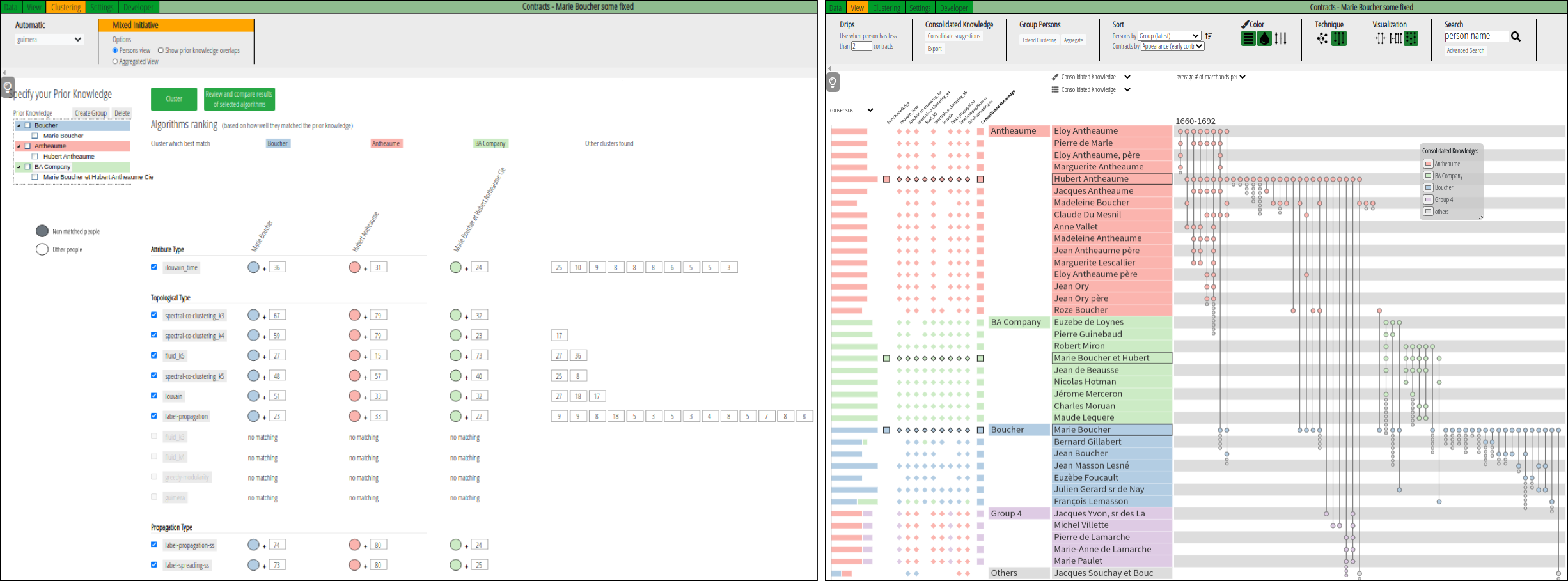}
  \caption{Two main phases of PK-clustering. On the left, the user has specified the Prior Knowledge (PK) groups (top left) and then reviews the list of algorithms ranked according to how well they match the PK. On the right, the user compared the detailed results of selected algorithms and consolidated the results. From the initial specification of three groups and three people, 4 relevant clusters were obtained with 37 people in total, plus one unclassified node (\textit{Others} group).}
 
	\label{fig:teaser}
}

\vgtcinsertpkg

\begin{document}


\firstsection{Introduction}

\maketitle

The goal of this work is to help social scientists, such as historians and sociologists, create meaningful clusters from social networks they study. 
In contrast to the belief that most data is easily available on the Web, as of today, most social scientists spend a long time collecting data, to construct social networks, based on documents or surveys,  
\begin{revs}in order to create and carefully validate medium-sized networks (50--500 vertices).
\end{revs}
Before the start of the cluster analysis a great deal of effort goes into analysing other data and gathering knowledge (which we call prior knowledge in the rest of the paper). Social scientists study in great details the network entities (most of the time people), and the social ties they weave together, as it is the unit brick with which they can make historical or social hypothesis and conclusions.  
When the network is small, less than 30--50 nodes, it is possible to remember most of the relations and persons and visualization directly helps to show groups, hubs, disconnected entities, outliers, and other interpretable motifs. When the network grows larger, with hundred entities or millions of them, it becomes impossible to perform the visual analysis only at the entity level. The graph has to be summarized, and typically social scientists want to organize it in social \emph{communities}.  A large number of algorithms are available today to compute \emph{clusters} of entities from a graph, with the assumption that the computed clusters represent faithfully the social communities. However, most social scientists are not familiar with all of the available algorithms and are challenged to choose which algorithm to run, with which parameters, and how to reconcile the computed clusters with their prior knowledge. Furthermore, the clusters computed by the algorithms do not always align with the concept of community from the social scientists.

Typically, social scientists select an analysis tool based on their familiarity with the tool and the level of local or online support they can access. Therefore, they most often use popular systems such as R~\cite{Rstat}, Gephi~\cite{gephi}, Python with NetworkX~\cite{networkx}, or Pajek~\cite{pajek}. 
To compute clusters, they follow a strained process: they select and run algorithms provided in the tool and then try to make sense of the results (see~\autoref{fig:traditionalpkprocess}). When they are not satisfied or unsure, they iteratively tweak the parameters of the algorithms at hand, run them again and hope to get results more aligned with their prior knowledge. This analysis process is unsatisfactory for three main reasons:
\begin{wrapfigure}{r}{0.5\columnwidth}
    \includegraphics[width=0.5\columnwidth]{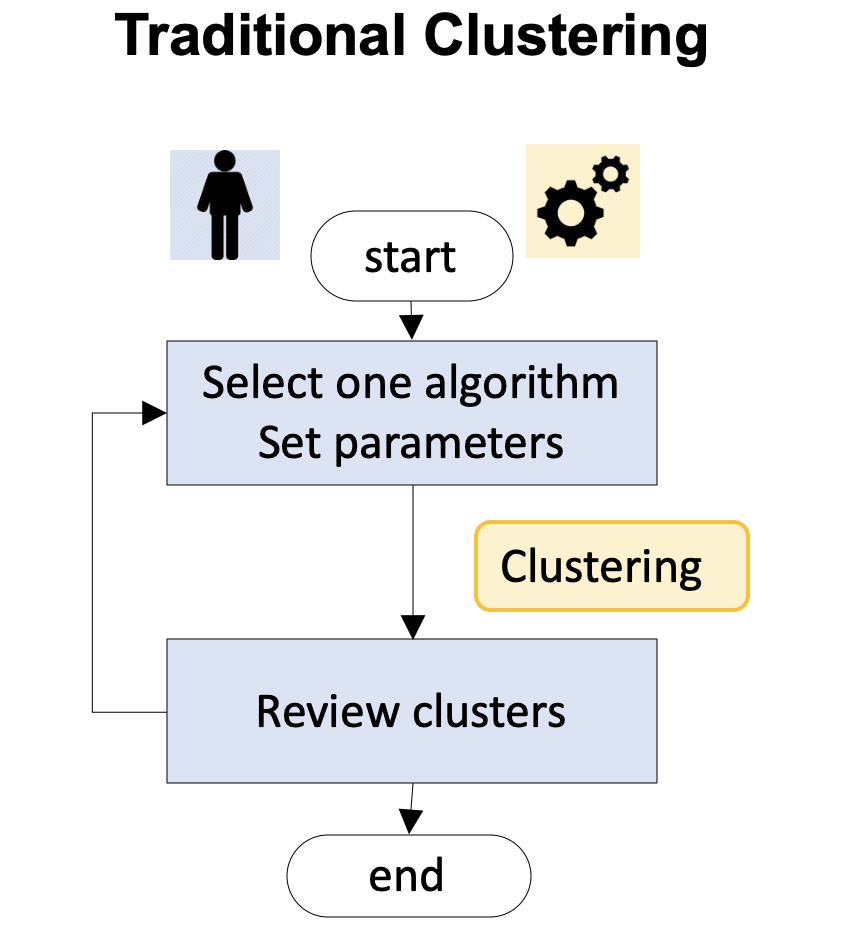}
    \caption{Traditional Clustering.  The output is a clustering, usually from a randomly chosen algorithm.}
    \label{fig:traditionalpkprocess}
\end{wrapfigure}
\begin{enumerate}[left=0pt .. \parindent,nosep]
\item it forces them to try a sometimes large number of black-box algorithms one by one, tweaking parameters that often do not make sense to them; \item even when a parameter makes sense to them, such as the number of clusters to compute, $k$ in $k$-means clustering, they have no clue of what value would generate good results, and are left with trial and error;
\item even if they could 
painstakingly evaluate the results of all clustering algorithms  according to their prior knowledge, no existing system allows users to do so easily, leading users to give up and blindly accept the results of one of the first algorithms they try.
\end{enumerate}
Those complaints have been heard repetitively during the decades our team has worked with social scientists.

Moreover, clustering is an ill-defined problem: for one dataset, there is no ground truth, and several partitions can be considered good according to the metric chosen to evaluate the result~\cite{kleinberg2003impossibility}. In a Social Sciences setting, this means, for example, that the same social network could be clustered to find families, friend groups, or business relationships. One partition is not better than the other: it depends on the purpose of the analysis. This problem increases the need for interactive tools, which lets the user specify which type of partition is expected.

To address those issues we propose a novel approach, called PK-clustering, which allows social scientists to iteratively construct and validate clusters using both their \emph{prior knowledge} and consensus among clustering algorithms. A prototype system illustrates such approach.


\begin{figure}
    \centering
    \includegraphics[width=\columnwidth]{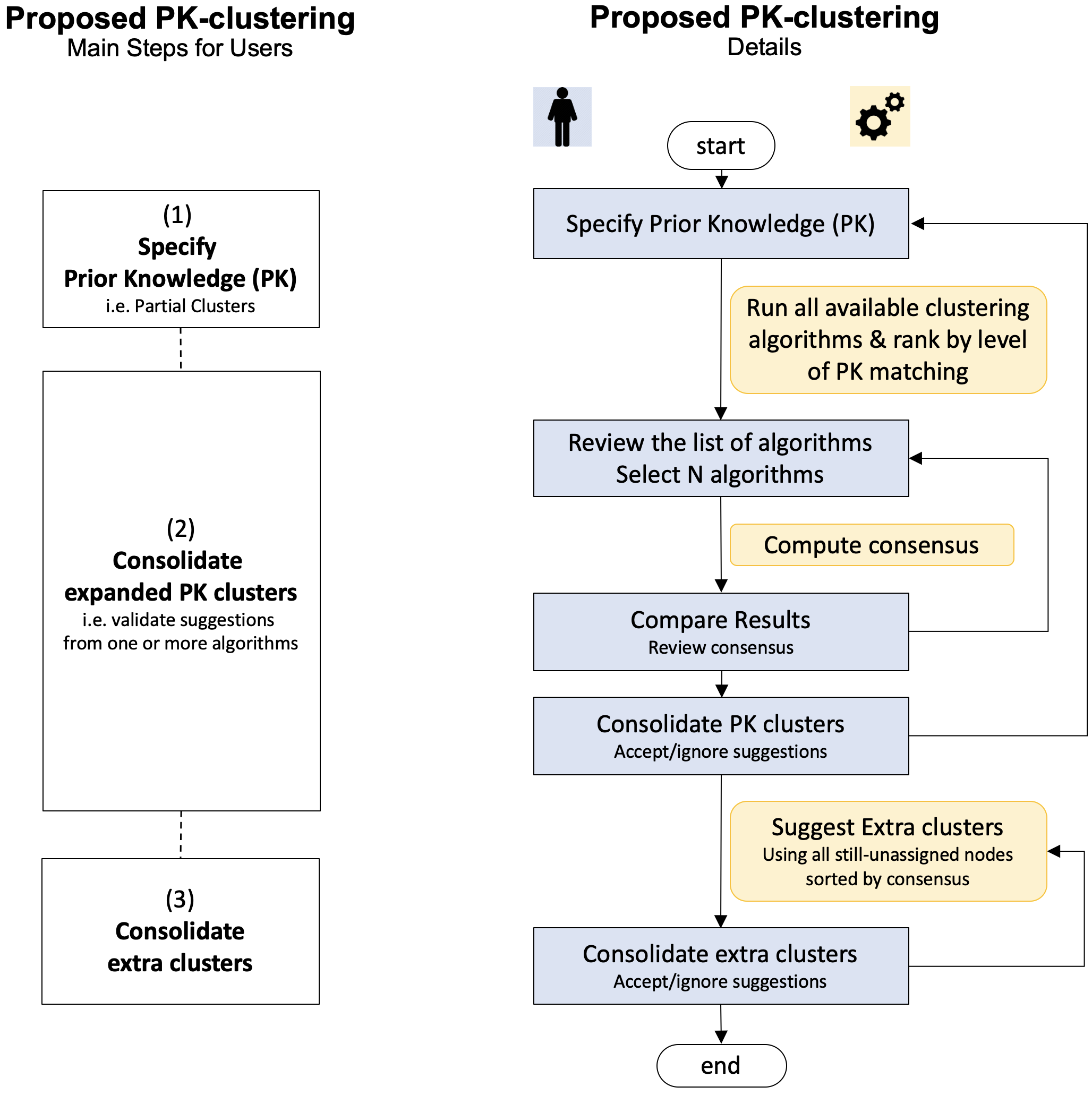}
    \caption{PK-clustering. The output is a clustering supported by algorithms and validated (fully or partially) according to the user's Prior Knowledge.}
    \label{fig:pkprocess}
\end{figure}

The proposed approach includes three main steps (see~\autoref{fig:pkprocess}):
\begin{enumerate}[nosep]
\item \textit{Specify Prior Knowledge (PK)}. Users  introduce their prior knowledge of the domain by defining partial clusters. The tool then runs all available clustering algorithms.
\item \textit{Consolidate expanded PK clusters}. Users review the list of algorithms, ranked according to how well they match the prior knowledge.  They compare results and consensus, then accept or ignore suggestions to expand the prior knowledge clusters
\item \textit{Consolidate extra clusters}. The tool suggests extra clusters on unassigned nodes. The user reviews consensus on each proposed cluster, then accepts or rejects suggestions.
\end{enumerate}

The output of the process is, using a direct quote from a social scientist providing feedback on the prototype: ``a clustering that is supported by algorithms and validated, fully or partially, by social scientists according to their prior knowledge".

\rev{According to the need to combine data mining with visualizations~\cite{Ben02DiscoveryTools} and inspired by the idea of letting the user collaborate with the machine to reach specific goals~\cite{Horvitz99}, the proposed approach follows a \rev{user-initiated} mixed-initiative~\cite{Horvitz99} visual analytics process.}

In our case, users focus on the results that expand on their prior knowledge, filter-out the most implausible results, but can readjust  when they realize that several algorithms are consensual despite not matching the prior knowledge (hinting at other possible meaningful structures). Our   mixed-initiative approach allows social scientists to seed the clustering process with a small set of well-known entities that will be quickly and robustly expanded into meaningful clusters (details in~\autoref{sec:pk-clustering}).

Contrary to a current trend\cite{molnar2019}, we do not aim to improve the interpretability of algorithms but to improve the interpretation of the results of black-box algorithms in light of prior knowledge, provided by the user.  Every day, we use complex mechanisms that we do not fully understand, like motorbikes, cars or electric vehicles using various kinds of engines, shifts, and gears, but we are still able to choose which one best fit our needs according to their external utility and not by understanding their complex internal machinery. In addition, it is usually more important to social scientists to find an algorithm that provides useful results than to understand why another algorithm failed to do so.

The main contributions of this article are:
\begin{enumerate}[left=0pt .. \parindent,nosep]
\item a new interactive clustering approach;
\item a prototype (shown in \autoref{fig:teaser}) implementing PK-clustering with 11 clustering algorithms of different families applied with different parameters configurations;
\item two case studies.
\end{enumerate}
\section{Related Work}

Our approach relies on several families of clustering methods and the visualization and exploration of their results. We first describe a brief overview of clustering for graphs, as well as semi-supervised methods, then several works in the literature related to \rev{visual analytics: interactive clustering, }groups in networks and ensemble cluster visualization.


\subsection{Graph Clustering}

One of the main properties of social networks is their community structure~\cite{Girvan7821} that reveals group relationships between nodes, known as communities or clusters, having higher density of edges than the rest of the graph. Similar characteristics or roles are often shared between nodes of the same community. 
In social networks, a community can mean a lot of things like families, workgroups, or friend groups. There is abundant and growing literature on clustering methods to find these communities for social networks. The majority of the research is made only on topological algorithms, \ie algorithms which use only the structure of the network to find clusters. \cite{FORTUNATO201075} proposes a description and a classification of various algorithms, such as divisive, spectral and dynamic algorithms, or methods, such as modularity-based, statistical inference, to cite a few.
\rev{In contrast, many multidimensional clustering algorithms use a distance function as parameter, but graph clustering algorithms mainly rely on the structure of the graph instead.}

Even if the majority of studies are based on simple graphs, real-word phenomena are often best modeled with bipartite graphs, also known as 2-mode networks. It is the case for social scientists, who often build their networks from raw documents containing mentions of people. In that case, it is more straightforward to model the persons as one set of nodes, the documents as the other one, and linking an individual to a document if the individual is mentioned in it. This is one of the reasons some research is made on bipartite graph community detection~\cite{alzahrani2016community}.

Moreover, recent new approaches try to use the attributes of the nodes~\cite{yang2013community} and the dynamic aspect of the networks~\cite{rossetiSurvey} to find more relevant communities. Some toolkits offer a large number of algorithms; for example, the Community Discovery Library (CDLIB)~\cite{cdlib} implements more than 30 clustering methods with variations inspired by 67 references.

\subsection{Semi-supervised Clustering}\label{sec:semisupervised}

In semi-supervised clustering the user integrates the data mining task with additional information to improve the clustering quality in terms of minimizing the error in assigning the cluster to each data of interest.

Semi-supervised clustering can be divided into constraint-based and seed-based clustering.
The former includes must-link (ML) and cannot-link (CL) constraints~\cite{basu08, wagstaff2001constrained}. ML($x,y$) indicates that given two items $x$ and $y$, they must belong to the same cluster, while CL($x,y$) means that $x$ and $y$ must belong to different clusters.

Seed-based clustering requires a small set of seeds to improve the clustering quality. Several works addressing seed-based clustering have been proposed in the literature, such as: $k$-means~\cite{basu02}, Fuzzy-CMeans~\cite{bensaid96}, hierarchical clustering~\cite{bohm08}, Density-Based Clustering~\cite{lelis09}, and graph-based clustering~\cite{wagstaff2001constrained}. 
Shang et al.~\cite{shang2017efficiently} use a seeding then expanding scheme to discover communities in a network. Their clustering method considers edges as documents and nodes as terms. 

Swant and Prabukumar~\cite{SAWANT2018} review graph-based semi-supervised learning methods in the domain of hyperspectral images. 
Nodes of the graph represent items that may be labeled, while the edges are used to specify the similarity among the items. 
The technique classifies unlabelled items according to the weighted distance from the labeled items.

\begin{revs}
\subsection{Mixed-Initiative Systems and Interactive Clustering}

Introduced by Horviz~\cite{Horvitz99}, mixed-initiative systems are ``interfaces that enable users and intelligent agents to collaborate efficiently''. Several Visual Analytics systems are based on mixed-initiative interactions, \eg~\cite{makonin16, cook15, zhou13, wall18}, in particular the interactive clustering systems.

PK-Clustering is an interactive clustering system. A review by Bae et al~\cite{baeetal20} shares our concerns:
``Real-world data may contain different plausible groupings, and a fully unsupervised clustering has no way to establish a grouping that suits the user’s needs, because this requires external domain knowledge.''
Interactive clustering systems aim at producing visual tools that let users interact and compare several clustering results with their parameter spaces, making it easier to find a satisfactory algorithm for a particular application. Several such systems exist (\eg~\cite{cavallo2018clustrophile, l2015xclusim}) but few deal with graph data.
These systems adapt one algorithm to become interactive using some type of constraints. 
Instead, our approach applies ML/CL constraints on a wide variety of existing algorithms, providing richer algorithms and control than the reviewed systems. 

\end{revs}




\subsection{Groups in Network Visualization}

\rev{
To assess the quality of clusters in graphs, the clusters should be visualized.
A state of the art report (STAR) on the visualization of group structures in graphs is proposed by Vehlow et al.~\cite{EVstar.groupstructures15}.}
Several strategies exist to display group information on top of node-link diagrams. Jianu et al. evaluated four of them: node coloring, LineSets, GMap and BubbleSets~\cite{Jianu14}. They show that BubbleSets is the best technique for tasks requiring group membership assessment.
But, displaying group information on a node-link diagram can reduce the accuracy by up to 25 percent when solving network tasks.
Another finding is that the use of GMap of prominent group labels improves memorability. Saket et al. evaluated the same four strategies~\cite{Saket14}, using new tasks assessing group-level understanding.

Holten~\cite{Holten2006HierEdgeBundles} proposes edge bundling on compound graphs. He bundles together adjacenct edges, making explicit group relationships at the cost of losing the detailed relationships.
A good example of manual grouping and tagging is SandBox, which allows users to organize bits of information and their provenance in order to conduct an analysis of competing hypotheses~\cite{proulx06sandbox}. \rev{A lot of work has also been done on the visualization of categorical variable in tabular data \cite{kosara2006parallel, gratzl2014domino}, which is similar to the notion of groups in networks}.


\subsection{Ensemble Clustering}
In the context of machine learning, an ensemble can be defined as ``a system that is constructed with a set of individual models working in parallel whose outputs are combined with a decision fusion strategy to produce a single answer for a given problem''~\cite{wang08}. Several strategies exist for combining multiple partitions of items in a clustering setting~\cite{strehl2002cluster}.
Concerning visualization research, Kumpf et al.~\cite{kumpf18} consider ensemble visualization as a sub-field of uncertainty visualization, for which some surveys exist~\cite{Bonneau2014, maceachren05}. They describe a novel interactive visual interface that shows the structural fluctuation of identified clusters, together with the discrepancy in cluster membership for specific instances and the incertitude in discovered trends of spatial locations.
They aim at identifying ensemble members that can be considered similar and propose three different compact representation of clustering memberships for each member.
\rev{Our system provides a consensus based interactive strategy that takes into account user's prior knowledge instead of relying on mathematically defined optimal assignments only.}

\subsection{Summary}\label{sec:summary}

The community detection problem in graphs has been studied in a lot of different settings. We can classify it this way from the user perspective:
\begin{description}[leftmargin=0pt,nosep]
    \item [Standard clustering.] One algorithm is picked with a set of parameters and the user check if the results are consistent with his prior knowledge, which is not represented in the process.
    \item [Ensemble clustering.] Many algorithms run with potentially many parameters, and a final partition is obtained by trying to merge optimally the partitions. At the end of the process, one clustering is given to the user who has to check if it is consistent with the prior knowledge, which is not used either.
    \item [Semi-supervised clustering.] The user provides the prior knowledge and lets the algorithm propose a final solution using this information in its computation. The results should be good by design, regarding the knowledge of the user.
\end{description}

The aim of our proposed framework is to combine these three approaches, to integrate the user in the analysis loop and allow him to have a better impact on the final community detection result.

\section{PK-clustering}

We present a new approach, inspired by the three types of clustering methods described in~\autoref{sec:summary}: Standard clustering, Ensemble clustering and Semi-supervised clustering. It runs a set of algorithms, then highlights those that best match the prior knowledge provided by the domain expert. 
The user then reviews and compares the results of the selected algorithms, in order to consolidate a satisfactory and consensual partition.

PK-clustering is not tied to any specific graph representation technique and could be used to augment any of them. Our prototype is implemented in the \paovis tool~\cite{paohvis} which illustrates how users can view their networks as PAOH (Parallel Aggregated Ordered Hypergraph) or traditional Node Link diagrams. PK-clustering relies heavily on having a list of nodes, so the PAOH representation is naturally adapted to PK-clustering, and will be used in all the figures. 
 
After a general overview of the process, we describe each step in more details, illustrated with screen samples taken during the analysis of a small fictitious network.

\subsection{Overview}
\label{sec:pk-clustering}

In PK-clustering the user and the system take turn to construct and validate clusters. 
The process involves three main steps, each with several activities (see ~\autoref{fig:pkprocess}.
The blue boxes describe the user activities while the yellow boxes describe the system activities.) 
After loading the dataset, the process is as follows:

\noindent \textbf{(1) Specify Prior Knowledge (PK)}. 
\begin{enumerate}[left=.3em,nosep,label={\arabic*}.]
\item The domain experts interactively specify the PK by defining  groups, \ie naming groups and assigning entities to them. 
Typically, an expert would assign a few items (1-3) to a few groups (2-5), thus creating a set of partial clusters. 
\item All available clustering algorithms are run. Algorithm parameters (\eg number of clusters) may also be varied manually or automatically using a grid search or a more sophisticated strategy, resulting in additional results. Depending on the type of algorithm, topology and/or data attributes are used. The specified PK is used by the semi-supervised algorithms, which are the only ones able to use it. 
\end{enumerate}

\noindent \textbf{(2) Consolidate expanded PK clusters}.
\begin{enumerate}[left=.3em,nosep,label={\arabic*}.,start=3]
\item Users review the ranked list of algorithms. They can see if the algorithm results match the PK completely, partially or not at all. Information about the number of clusters generated by each algorithm is also provided.  Users select the set of $N$ algorithms they think are the most appropriate. 
\item The consensus between the selected algorithms is computed and visualized next to the graph visualization  (in the \paovis display in our prototype)
\item Users review and compare the suggestions made by the algorithms to expand the PK-groups into larger clusters and examine consensus between algorithms.
\item Users accept, ignore, or change the cluster assignments. This consolidation phase is crucial, as users take into account their knowledge of the data, the graph visualization, and the results of the clustering algorithms to make their choices. 
\end{enumerate}

\noindent \textbf{(3) Consolidate extra clusters}. 
\begin{enumerate}[left=.3em,nosep,label={\arabic*}.,start=7,itemindent=0pt]
\item The system proposes extra clusters using nodes that have not being consolidated yet and remain unassigned. Users can select any algorithm and see the extra clusters it suggests.  
\item For each proposed cluster, users can see if other algorithms have found similar clusters, and then consolidate again by accepting, ignoring, or changing the suggestions for all the nodes in the proposed cluster.  This step is repeated with other clusters until the user is satisfied.
\end{enumerate}

/dev{At any point users can go back, select different algorithms, or even change the PK specification to add new partial clusters.  Users can also opt not to specify any PK partial clusters at all, and accept all consensual suggestions without reviewing them in details. This gives users control over how much they want to be involved in the process. Similarly, users are not required to assign every single node to a cluster.}

\begin{wrapfigure}{r}{0.4\linewidth}
    \includegraphics[trim=15 10 10 10,width=\linewidth]{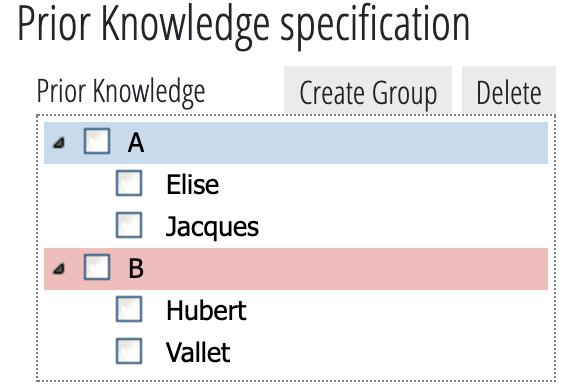}      
    \caption{Prior Knowledge specification, the user defined two groups composed of two members.}
    \label{fig:Small-PK_specification}
\end{wrapfigure}
By specifying the PK in the first phase, before running the algorithms, users avoid being influenced by the first clustering results they encounter.  The process leads to algorithms whose results match the PK, but it also allows to review results that contradict it.

We believe that PK-clustering addresses the important problems identified in the introduction: it helps users decide which algorithm(s) to use, facilitates the review of the results taking into consideration both the consensus between algorithms and the knowledge users have of their data.
We will now review each step in more details.


\subsection{Specification of Prior Knowledge}

We ask users to represent prior knowledge as a set of groups. Each group contains the node(s) that the expert is confident belong to the defined group.
In the case of \autoref{fig:Small-PK_specification}, each of the two prior knowledge groups contains two nodes, and it specifies that the user is expecting to see at least two clusters, with the first two people in a blue cluster A, and the other two in a red cluster B.
This representation expresses \emph{must-link} and \emph{cannot-link} constraints described in \autoref{sec:semisupervised} in a simple visual and compact form. It is not required to specify all binary constraints because the information is derived from the prior knowledge groups. 

\subsection{Running the Clustering Algorithms} \label{sub:families}

Our prototype includes 11 algorithms taken from three families:
\begin{description}[leftmargin=0pt,nosep]
\item[Attribute based algorithms.] Graph nodes can have intrinsic or computed attributes that can be used for grouping, such as gender, family name and age. Some community detection algorithms use those attributes alone or together with the topology to partition the graph. 
A clustering algorithm considers attributes according to their type. For categorical attributes (\eg male / female) it finds matching attributes and merges them if necessary. For numerical attributes (\eg income) the algorithm seeks to define intervals which can be adjusted for propagating clusters.
Algorithms in this family can also use multiple attributes together.

\item[Topology based algorithms.] Most of the clustering algorithms consider only the graph topology \cite{baroni17} and try to optimize a topological measure such as \emph{modularity} \cite{brandes08}.
Those algorithms only use the connections between the people to find groups. Their aim is to find groups of nodes such that the density of edges is higher between the nodes of one group than between the group and the rest of the graph.

\item[Propagation / Learning based algorithms.] Semi-supervised machine learning algorithms learn from an incomplete labeling of data and use it to classify the rest of the data. They represent a class of machine learning methods, also called label propagation methods, which can take into account users' Prior Knowledge groups in its clusters computation. 
By design, this type of algorithms will always provide a perfect match with the Prior Knowledge, even if the Prior Knowledge makes no sense.

\end{description}









Our prototype implements 2 attribute based algorithms (one for numerical attributes and the other for categorical attributes), 7 topology based algorithms and 2 propagation based. Since we often deal with hypergraphs 2 of the topology-based algorithms are bipartite node clustering algorithms: Spectral-co-Clustering \cite{coClustering} and Bipartite Modularity Optimisation. Since the majority of community detection algorithms are for unipartite graphs, we perform a projection into a one-mode network \cite{bipartiteProjection}. Basically, each pair of nodes which are in the same hyperedge are connected together in the resulting graph, with a weight being the number of shared hyperedges~\cite{guimera2007module}.

\begin{revs}
Some algorithms require parameters to be specified. We do not force the user to specify values for all the parameters, when possible, we infer them from the PK-groups. For instance, instead of using an arbitrary default for the number of expected clusters $k$ in $k$-means clustering, we run the algorithm several times with a value of $k$ from the number of specified PK-groups to this number plus two. Therefore, our implementation computes a total of 15 clustering algorithms ($11+4$). 
The strategy of using several parameter combinations for the same algorithm is often used in ensemble clustering to increase the number of different clusterings.
However, the number of parameter combinations can be extremely high. The research field of \emph{visual parameter space exploration} (see \eg \cite{6876043}) is devoted to exploring this space of parameter values in a sensible way; we currently address the problem only for simple cases. 


Once all the algorithms finish the computation, we try to match the resulting partitions with the PK and rank the algorithms by how interesting their results are likely to be for the user.
\end{revs}


\subsection{Matching Clustering Results and Prior Knowledge}
\label{sec:matching}

Once a clustering is computed, we want to know how well it is compatible to the PK, and if possible, match every PK-group with a specific cluster. 
We use the \textit{edit distance} to measure this matching, as its computation allows us to directly link each PK-group to a specific cluster. 
Given two partitions, the edit distance is the number of single transitions to transform the first partition into the second one. 
For example, the edit distance between the two partitions of 4 nodes $ P_1 = \{\{1,2,3\}, \{4\}\}$ and  $ P_2 = \{\{1,2\}, \{3,4\}\}$ is 1 because moving the node 3 from the first to the second set of $P_1$ would transform it into $P_2$. 
A clustering can be seen as a partition since every node has a label, but the PK can only be seen as a partial partition because only some nodes are labeled. 
We say that the edit distance between the PK and a clustering is 0 if every group of the PK is a subset of an exclusive cluster, \ie if every person of a PK-group is retrieved in the same cluster, with no overlaps. 
Thus, we define the edit distance as the number of node transitions between the groups of the PK to get to the state where each group is a subset of an exclusive cluster. 


To compute the edit distance and the matching, we build a bipartite graph: each meta-node corresponds either to a PK-group, or a cluster. We then link them if they share a node, with a weight equals to the number of shared nodes.
Computing the edit distance and producing a matching between the PK-groups and the clusters is then equivalent to the assignment problem, where the goal is to find a maximum-weight matching in the graph. \cite{Assignment}.

Once this matching is computed, the total sum of the weights minus the sum of the weight of the matching is equivalent to the number of transitions needed to transform the first partition into the second one (or the PK into a sub-partition where each set is an exclusive subset of the sets of the second partition), \ie the edit distance.

\begin{wrapfigure}{r}{0.5\linewidth}
    \centering
    \includegraphics[width=\linewidth]{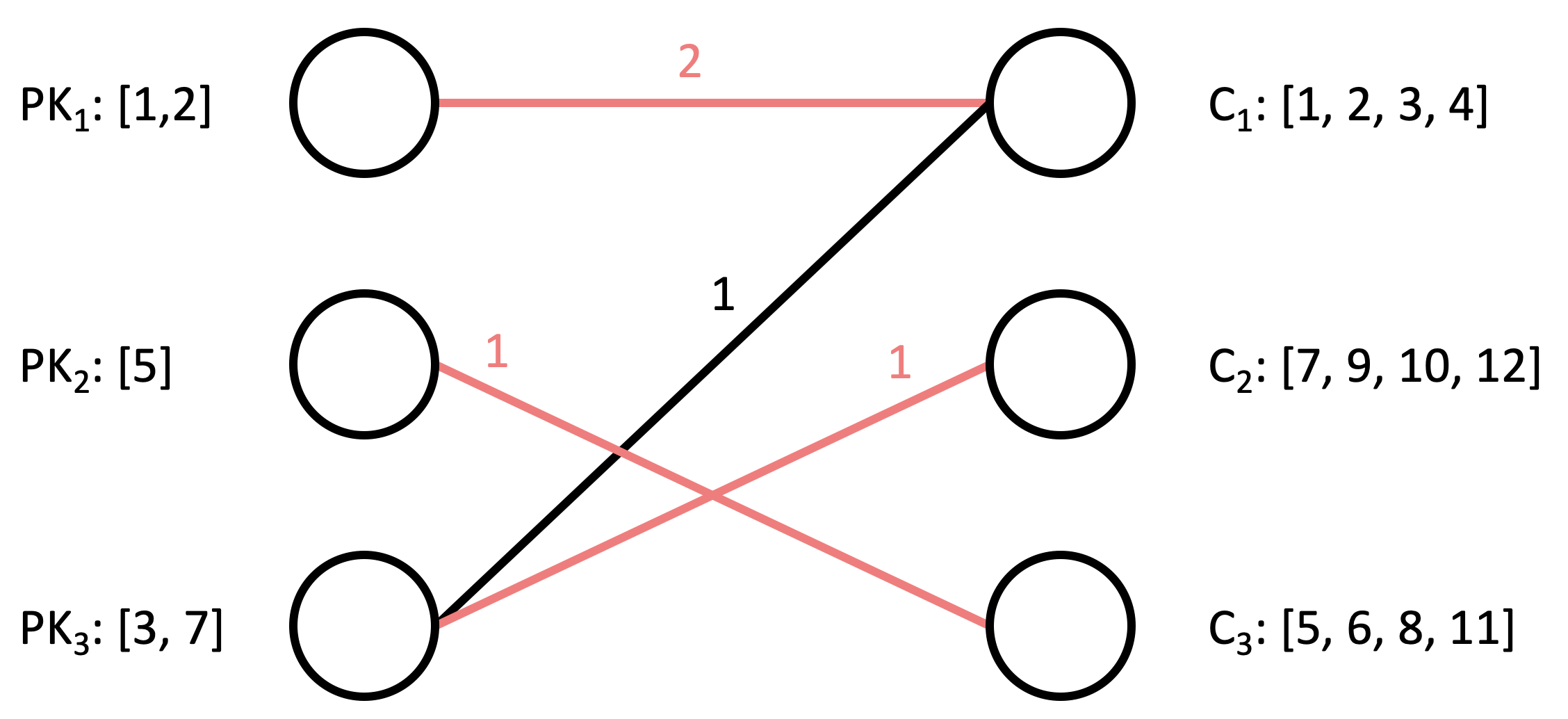}    
    \caption{Red edges represent the prior knowledge matching}
    \label{fig:BipartiteMatching}
\end{wrapfigure}
For example, given a clustering of 12 nodes $N = {1,2, \ldots, 12}$, the clusters $C_1 = [1,2,3,4]$, $C_2 = [7,9,10,12]$ and $C_3 = [5,6,8,11]$ and a PK composed of 3 groups $PK_1 = [1,2]$, $PK_2 = [5]$ and $PK_3 = [3,7]$, the maximum-weight matching is given by the edges $(PK_1, C_1)$, $(PK_2, C_3)$ and $(PK_3, C_2)$. This is illustrated in ~\autoref{fig:BipartiteMatching}. The edges of the matching correspond to the matching between the PK-groups and the clusters. The edit distance is then equal to the sum of all the weights of the bipartite graph minus the sum of the weights of the maximum matching (in red), thus equaling $5 - 4 = 1$. In other words, we only have to move the node 3 from $PK_3$ to $PK_1$, for every PK-group to be a subset of an unique cluster, with no overlap.



In the end, we hope to find matches linking every PK-group to one specific cluster, with no overlaps. This is not always the case and sometimes two or more PK-groups are subsets of the same cluster. In that case, it is not possible to link all these PK-groups to the same cluster since we want one unique cluster for each group. Thus, we say that the algorithm failed to match the prior knowledge and we do not summarize it visually.  




\subsection{Ranking the Algorithms}

\rev{
The algorithms are ranked by their degree of matching with the prior knowledge, using the edit distance. We also introduce a \emph{parsimony} criterion if there is a tie between two or more algorithms. The algorithm with the smaller number of other clusters will be shown first, as the results are easier to interpret. Moreover, the number of specified prior knowledge groups is expected to be close to the final number of clusters the user wants to retrieve, as social scientists often have a good knowledge of their data.
}

To complement the parsimony rule, we also consider that the family of propagation/learning based clustering algorithms is more complex than the two previous families (attribute or topological based clustering), \rev{in the sense that they are more difficult to explain.}
If a simple and a complex algorithm match the prior knowledge, the simpler one is presented first. For example, if grouping by the attribute ``profession'' provides a perfect match, then it is ranked higher than a propagation based method achieving the same perfect match.

\rev{
Semi-supervised methods will always provide a perfect match by definition. But if all the other algorithms (topological and attribute based) do not give a match, it means that the PK does not align well with the data. This would signal the user to reconsider his PK or provides more information in the graph.}

\subsection{Reviewing the Ranked List of Algorithms}


Once the clustering algorithms have been matched with the PK, users can review the list of algorithms, ranked by how well their results match the PK.
\autoref{fig:matchingvis} shows two modalities to visualize the ranked list (individual nodes, and aggregate representation). We will describe in details the first modality, which shows individual nodes as small colored circles (also used on the left of~\autoref{fig:teaser}):

\begin{wrapfigure}{r}{0.5\linewidth}
 \centering
    \includegraphics[width=\linewidth]{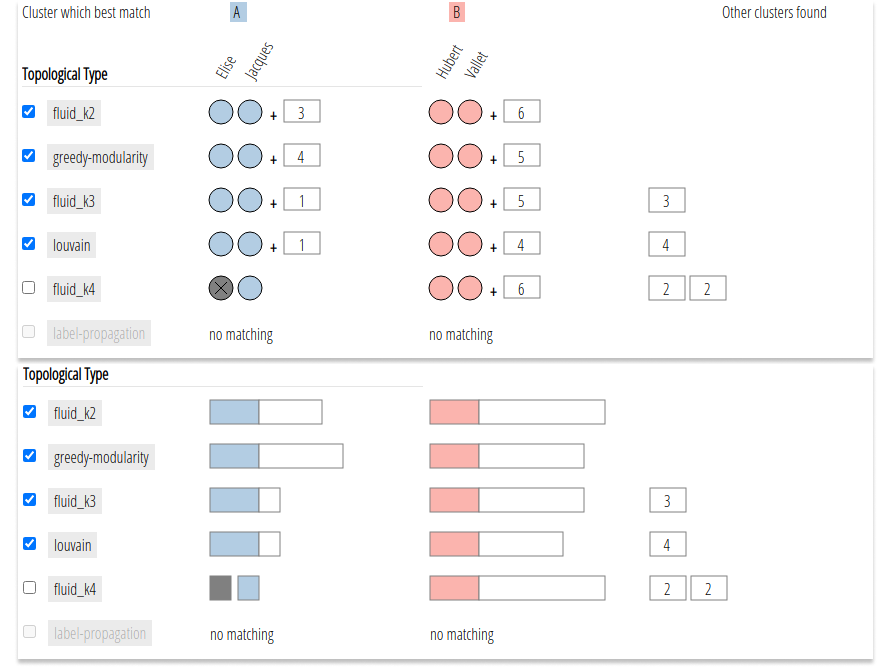}
    \caption{Two different modalities for the ranked list of algorithms. Top: persons are shown as circles. Bottom: aggregated view. Colors indicate the matching group. Gray indicates no match. White indicates extra nodes or clusters.}
    \label{fig:matchingvis}
\end{wrapfigure}

Each row is an algorithm, and the algorithms are grouped by family. On the right of the name of the algorithm we can see a representation of the clusters that best match each of the PK-groups. In~\autoref{fig:matchingvis} we first see the cluster which best matches the blue PK-group, and then the cluster which best matches the red PK-group.  In each cluster we see colored dots for each person that matches, and dark gray dots with a X for no match. Additional nodes in the cluster are represented as white dots with a number next to it.  On the right most we see how many other clusters (if any) have been found by the algorithm - also represented as white dots with a number next to it.  

So for example, the second algorithm \emph{fluid\_k3} has a blue cluster that matches the blue PK-group plus 1 extra node, a red cluster that matches the red PK-group plus 5 nodes, and one extra cluster.  We see that the top four algorithms match the PK perfectly, while the next one \emph{fluid\_k4}  have a partial match. At the bottom, an algorithm has no match.

The alternate modality of representing the matches (shown at the bottom of~\autoref{fig:matchingvis}) uses bars to aggregate the nodes and show the proportion of matching, non-matching and other nodes in each cluster . This is more useful when dealing with bigger graphs, because it allows the user to see the results in a more compact way.






Once users have reviewed the list of algorithms they can review results of a single algorithm, or review and compare the results of all the selected algorithms.
By default only the top algorithms are selected for inspection, but users can select any set of algorithms according to different criterion: 
the \textit{degree of matching} (\ie they can choose to look at algorithms with no match to challenge their prior knowledge);
the \textit{algorithm type} (the user may prefer an attribute-based algorithm, rather than one based on topology); 
the \textit{size} of the matched clusters; 
or the number and size of \textit{other clusters} found by the algorithm. 

\begin{revs}
PK-Clustering expresses its prior knowledge through \emph{must-link} and \emph{cannot-link} constraints. However, at this stage, the user can decide to use this expressive power as strong constraints---only selecting algorithms that match all of them---or as weak constraints---to explore clustering results that support most or some of them. Our historian colleagues have used both, either to cluster a well-understood dataset with strong constraints or to generate hypotheses on less known ones.
\end{revs}


\rev{\subsection{Reviewing and Consolidating Final Results }
To consolidate the final results several approaches are possible. Applying mixed-initiative principles users can rapidly accept labels from a specific algorithm (which is particularly useful for large datasets), or review consensus between selected algorithms then accept only consensual suggestions, or dig in manually to review labels one by one, override labels when appropriate, or leave certain nodes unlabeled.  The tool generally  guides users to first focus on the PK clusters, then other clusters. 
The notion of prior knowledge can evolve during the exploration and the process can be iterated from the beginning when new knowledge is gained, thus giving new algorithm matches. Therefore, the approach is not linear but can be iterative.}

\subsubsection{Reviewing Results of a Single Algorithm}
By clicking on an algorithm name the results of that algorithm are displayed in the \paovis view (see~\autoref{fig:initialView}). \rev{In this view, each line corresponds to a person in the graph, and each vertical line represents an hyperedge connecting them~\cite{paohvis}, in a way visually similar to the UpSet representation~\cite{lex2014upset} but semantically different.  Alternative graph representations are available as well---such as node link diagrams---but the \paovis view is well adapted to PK-Clustering.}


Names are grouped by the proposed clusters. Clusters that match the prior knowledge are at the top, colored by their respective colors.  Black borders around labels highlight nodes that belong to the PK, making them easy to find.   All the other (non PK) clusters are initially regrouped in a single group labeled \textit{Others}.
A click on the \textit{Others} label expands the group into the additional clusters defined by the selected algorithm. 
Users can rename the clusters, and change which algorithm is used for grouping and coloring the nodes.

\subsubsection{Comparing Multiple Algorithm Results}

\begin{figure}
    \centering
    \includegraphics[width=\linewidth]{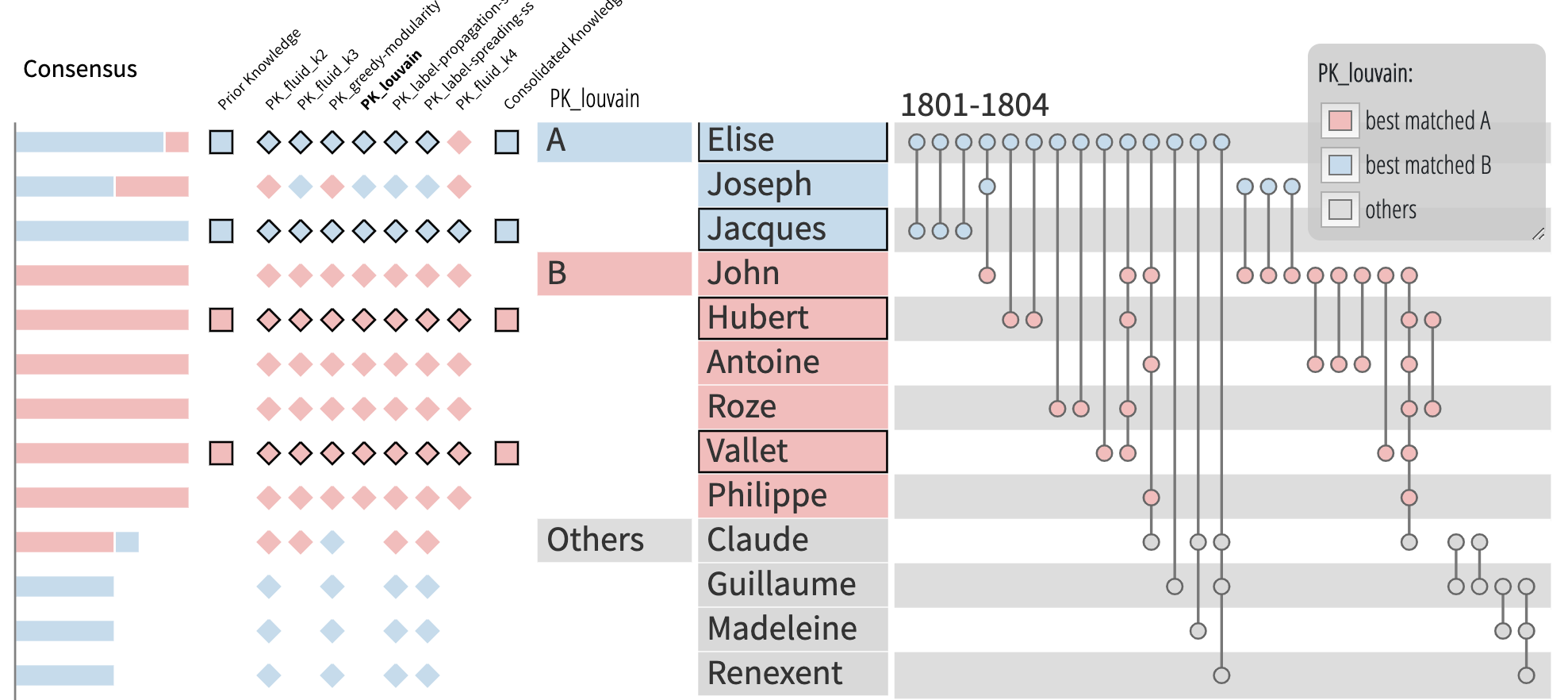}
    \caption{Reviewing and comparing results of multiple algorithms. One algorithm is selected to order the names and group them, but icons show how other algorithms cluster the nodes differently, summarized in the consensus bar on the left.}
    \label{fig:initialView}
\end{figure}

From the ranked list of algorithms users can select a set of algorithms and click the large green button to review and compare the selected algorithms in the \paovis view (see ~\autoref{fig:initialView}
and also ~\autoref{fig:teaser} for overall context). By default, the \paovis view groups the names using the clusters of the 1st algorithm, but on the left of the node names now appears complementary information about the results of all the selected algorithms.

On the far left, the consensus distribution appears as a horizontal stacked bar chart. The size of bar segments corresponds to the number of algorithms that associate the specific node to the cluster having the same color. On the right of the stacked bar chart, first appears the prior knowledge (with square icons). Icons and names of PK nodes have a black border. Further right are shown the individual algorithms' results, represented by diamonds, one for each node and algorithm. When the node is classified in one of the clusters matching a PK-group the diamond is colored with the color of that group.

For each node, the horizontal pattern of colored diamonds quickly tell users if there is agreement among the algorithms. If all algorithms agree the line of diamonds is of a single color. Conversely, if they disagree diamonds will vary in color. If a node does not match any PK-group then no icon is displayed in this phase.

In ~\autoref{fig:initialView} PK\_louvain is selected as the base algorithm for the grouping of names in the list. We see that there is very good consensus on the red cluster, but in the blue cluster only 4 out of 7 algorithms see Joseph as belonging to it.  Others see him as belonging to the red cluster. In \emph{Others}, 4 algorithms consistently disagree by assigning 3 more nodes to the blue cluster.
There are clearly many ways to cluster data, and users must decide the more meaningful one, based on their deep knowledge of the people in the network before validating clusters, possibly by re-reading source documents or gathering more.

\begin{figure}
    \centering
    \includegraphics[width=\linewidth]{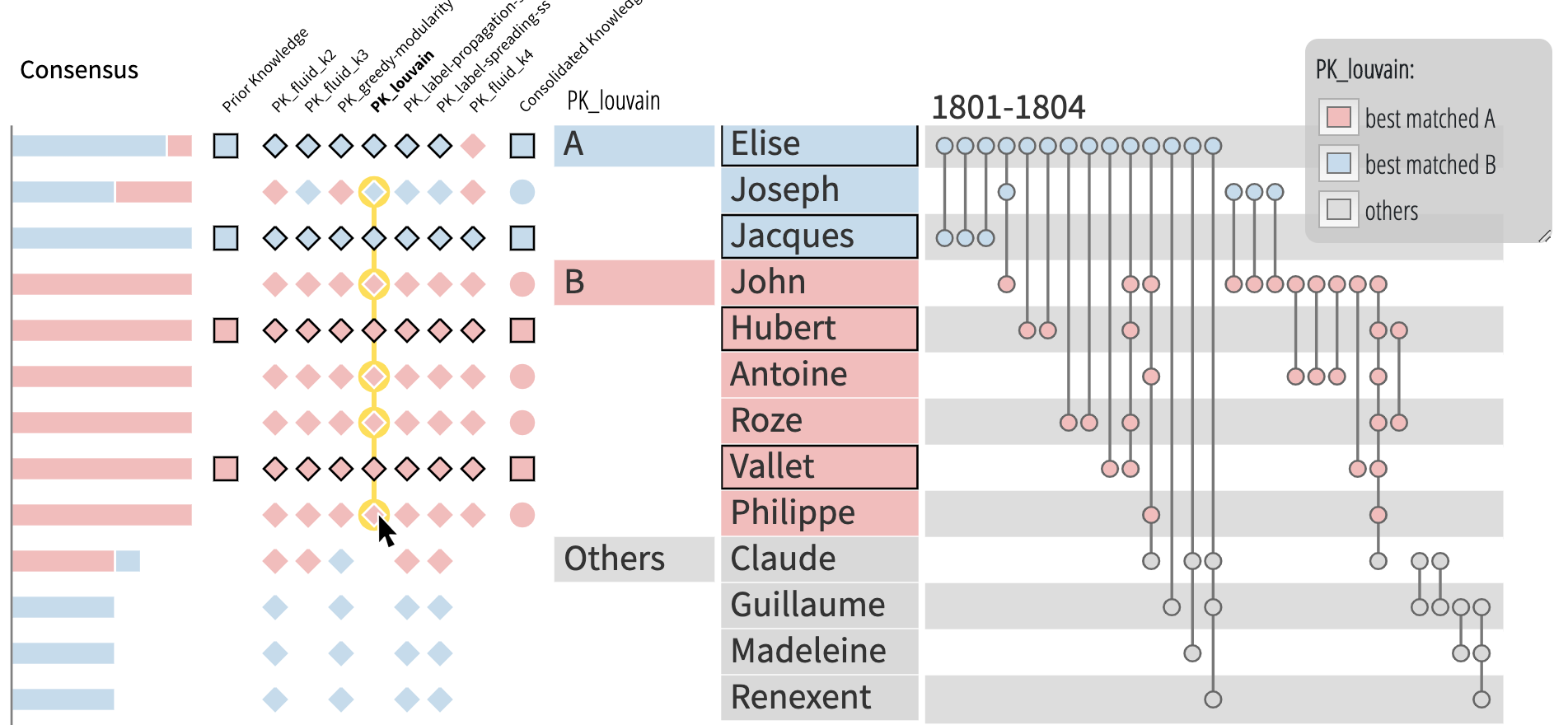}    
    \caption{The user quickly drags on consecutive icons (in yellow) representing the suggestions made by one algorithm to validate node clustering. Once the cursor is released the validated nodes appear as squares icons in the Consolidated Knowledge column.}
    \label{fig:multipleValidation}
\end{figure}

\subsubsection{Consolidating the prior knowledge clusters}
\label{sec:validating-pk}

Next, using their knowledge and the consensus of the algorithms, users validate clusters that expand the prior knowledge groups. We call the validated data \emph{consolidated knowledge}. It is kept in an additional column on the right of the algorithms, left of the names. The tool provides several ways to consolidate knowledge \rev{and keeps track of the decisions}:

\begin{description}[leftmargin=0pt,nosep]
        
    \item [Partial Copy.] By clicking on one of the icons or dragging the cursor down on a set of icons, users validate the suggestion(s) of an algorithm, adding colored squares in the consolidation column. Once this validation is done, the squares do not change color anymore and represent the user's final decision (unless changed manually again). \autoref{fig:multipleValidation} shows how a user drag-selects a set of diamonds in the column PK\_fluid\_k4. They are connected by a yellow line, which appears while dragging over the icons. 
    When done the status of the nodes in the Consolidated Knowledge column (rightmost) will change to square. 
    
    \item [Consensus slider.] Users can set the consensus slider to a certain value (for example 4) to automatically select all nodes that have been classified in the same cluster by at least 4 algorithms. While the slider is being manipulated circles appear in the consolidated column.  Then users can validate the suggestions \rev{by clicking or dragging on the circles, or by using the \emph{consolidate suggestions} button which will validate all suggestions at once. This button is shown in Fig. \ref{fig:teaser}.}  In summary, diamonds represent suggestions from one algorithm, circles temporary choices, and squares represents the knowledge validated by the user.


    
    \item [Direct tagging.] At any time, users can manually overwrite the association of a node to a cluster by right clicking on the node in the consolidated knowledge column and selecting an cluster from a menu.
    When no clear decision can be made users can leave nodes unassigned, and no shape is displayed in the consolidated knowledge column.
\end{description}


\subsubsection{Consolidating extra clusters}

The last step of PK-clustering aims to find new clusters for the nodes that have not been validated yet, based on the consensus of the selected algorithms. The suggestions are made from the point of view of one clustering algorithm that the user can change along the process. 
First, the user selects one algorithm in the PAOHVis view and the nodes are grouped by the clusters found by the algorithm. 
iThe PK-clusters are displayed at the top, followed by \emph{Others}, which contains everyone else. When users click on \emph{Others}, the other clusters are displayed ordered by consensus. Since the number of clusters can be high, all new clusters appear in gray to avoid the rainbow effect. A secondary matching process matches the clusters of the current algorithm with those of all the other algorithms, one by one (similar to the matching process described in~\autoref{sec:matching}) . Once the matching is done, the consensus of one cluster is computed as the sum of the cardinalities of the intersections between the cluster and all the other clusters of the other algorithms matched with it, divided by the number of nodes of the cluster.

\begin{figure}
    \centering
    \includegraphics[width=\linewidth]{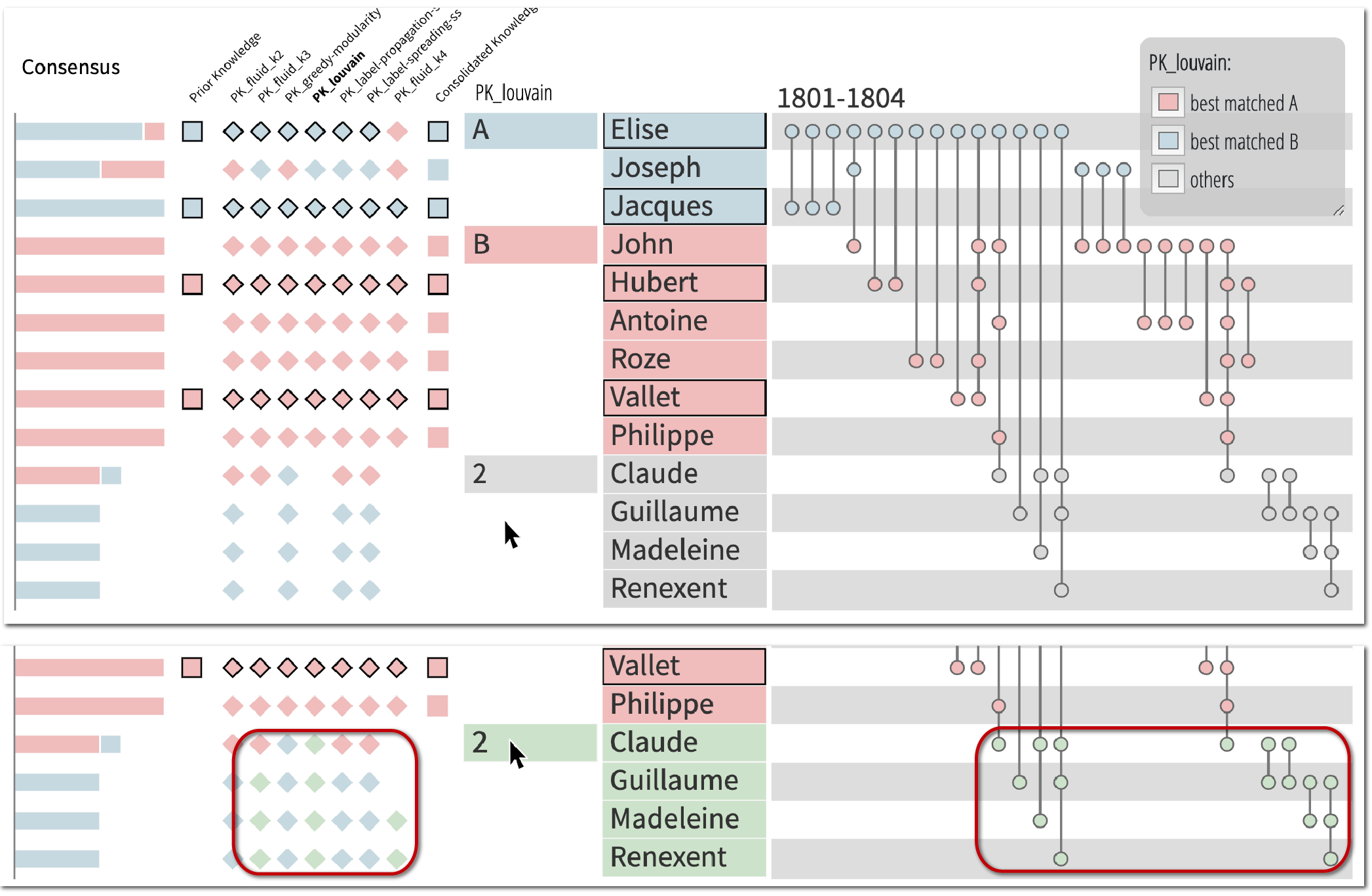}    
    \caption{Suggestion of extra clusters. The two PK-groups (red and blue) are validated (nodes in the consensus column are all squared). One extra clusters is proposed by the Louvain algorithm, labeled as 2. Hovering over the cluster 2, the consensus is displayed by the green diamonds. This feedback is also visible in the graph.}
    \label{fig:extraClusteringInteraction}
\end{figure}

When users hover over one cluster name, a new color is given to that cluster (\eg green) and new (green) diamonds appear for each algorithm that match the cluster and for each node that is assigned to the cluster (\autoref{fig:extraClusteringInteraction}).
Users can therefore see if the selected cluster is consensual, and with which algorithms.  The top part of~\autoref{fig:extraClusteringInteraction} shows the mouse pointer before hovering on the cluster 2. The bottom part shows that hovering the mouse pointer over the cluster 2, it changes to green and several green diamonds appear along three columns.

    
\begin{figure}
    \centering
    \includegraphics[width=\linewidth]{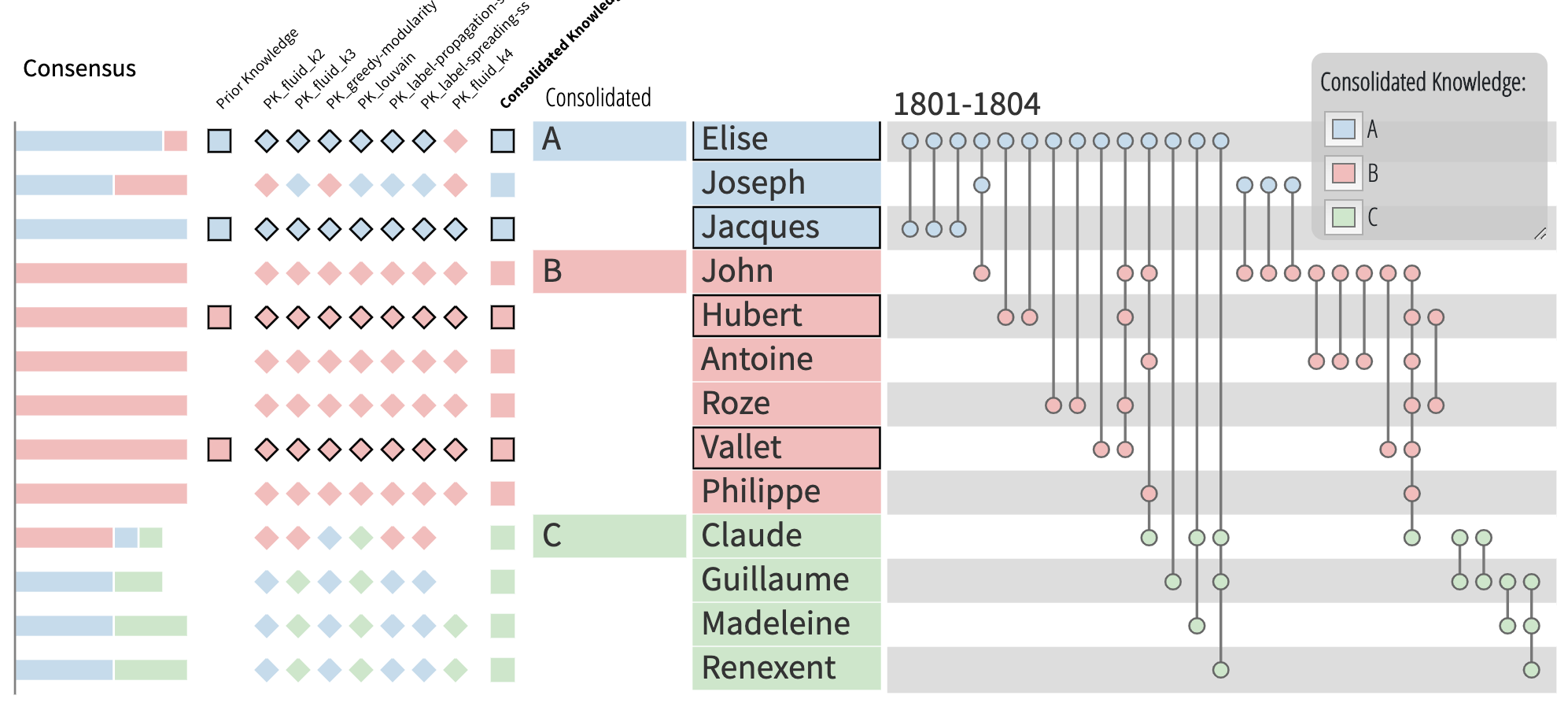}    
    \caption{The dataset has been fully consolidated. The persons are grouped and colored by the consolidated knowledge. The user decided to assign Claude, Guillaume, Madeleine and Renexent to cluster $C$, by taking into account the graph and the consensus of the algorithms.}
    \label{fig:extendedPartialConsolidatedDataset}
\end{figure}

The evaluation of the best cluster for a node can be done using multiple encodings. 
The suggested clusters appear into the consensus bar chart, in the set of algorithm output and when hovering over the node. A click on the color will validate the node into the cluster having that color.
If users are satisfied with the association proposed by the current algorithm, they can validate it by clicking on the cluster name. This will create a new group, so the user can classify the nodes into this new group, as seen before (\autoref{sec:validating-pk}): using the consensus slider, copying an algorithm result, or through manual labeling.
This process is repeated for the other clusters until there are no unlabeled nodes or the user is satisfied with the partial clustering. An example of a fully consolidated dataset is shown in \autoref{fig:extendedPartialConsolidatedDataset}.




\begin{revs}
\subsection {Wrapping up and Reporting Results}

At any stage of the process, the user can finish instantaneously, either by not labeling undecided nodes, or selecting and validating the results of a single algorithm---as traditional approaches do, or by using a specified threshold of consensus and not labeling the remaining entities. The appropriateness of the choice is up to the user and should be documented in the publication.

In addition to the consolidated clustering, the output of PK-clustering consists of provenance information in the form of a table and a summary report.
The table provides, for each vertex, the consolidated label, along with the labels produced by all the selected algorithms, and a description of the interaction that has led to the consolidation, such as ``selected from \verb|algorithm x|'', ``consensus $>= 5$'', or ``override'' when manually selected by the user instead of selected from an algorithm.  The summary provides counts of how many nodes were labeled using the different interactions methods and can be used in a publication. Examples are provided in the Supplemental Materials (as Fig. 2 and Fig. 5). 

Clustering results can thus be reviewed in a more transparent manner, revealing the decisions taken. In contrast, traditional reporting in the Humanities rarely questions or discusses how choices were made and merely mentions the  algorithm and parameters used. 

\end{revs}

\section{Case studies}

We describe two case studies using realistic scenarios where the clustering has no ground truth solution but has consequences, scientific or practical. We also report on the feedback received from practitioners.

\subsection{Marie Boucher Social Network}
\label{sud:MB}

We asked our historian colleague her prior knowledge on her network about the trades of Marie Boucher~\cite{Dufournaud17},  
composed of two main families: Antheaume and Boucher. Family ties were important for merchants, but could not scale above a certain level. Marie Boucher expanded her trade network far beyond that limit. She then had to connect to bankers, investors, and foreign traders, far outside her family and yet connected to it indirectly. 
As hinted in her article, Dufournaud believes that the network can be split in three clusters: one related to the Boucher family, one to the Antheaume family, and the third to the Boucher \& Antheaume company. Using standard visualization tools, she could see different connection patterns over time, but she wanted to validate her hypothesis using more formal measures and computational methods.



So she specified her hypotheses as Prior Knowledge and started the analysis. \autoref{fig:teaser} (top left) shows the three PK groups: Marie Boucher for the Boucher family, Hubert Antheaume for the Antheaume family, and the Boucher \& Antheaume corporation alone for the company. 

After running the algorithms, 9 algorithms produced a perfect match out of the 13 executed (see \autoref{fig:teaser} - left.) with the first algorithm listed an attribute based algorithm that uses the time attribute in its computation. That summary alone was found very interesting because the 3 clusters seemed very consensual among all the 9 algorithms, and furthermore, they appeared explainable by time alone..

In the PAOH view, she started by consolidating the 3 PK-groups using the amount of consensus among the algorithms as well as the graph representation and her own knowledge of the persons. At the end of this step, the Boucher, Antheaume, and Boucher \& Antheaume groups were consolidated, but there were still several persons not labeled on the consolidated knowledge. She decided to review in more detail the clustering results using the \emph{ilouvain\_time} algorithm because of its reliance on the time attribute, and also because its results seemed good in the matching view. After clicking on the virtual group \emph{Others}, the four other clusters computed by \emph{ilouvain\_time} appeared and were reviewed  by hovering the mouse on the names of these new groups. She selected only one clusters she was comfident about and consolidated it.

The final validated partition of the dataset is represented in \autoref{fig:teaser} (right).
The persons are colored and grouped by the consolidated knowledge. We can see that the final grouping makes sense in the PAOH visualization on the right. Only one person is not part of any group: Jacques Souchay. It is not unusual in historical sources to have persons mentioned without any information on them. 

Our historian colleague can now publish a follow-up article validating her hypotheses. \rev{The summary report will help document where the final grouping came from, increasing trust with regard to her claims.}

\subsection{Lineages at VAST}
\label{sud:VAST}

\begin{figure}[tb]
    \centering
    \includegraphics[width=\linewidth]{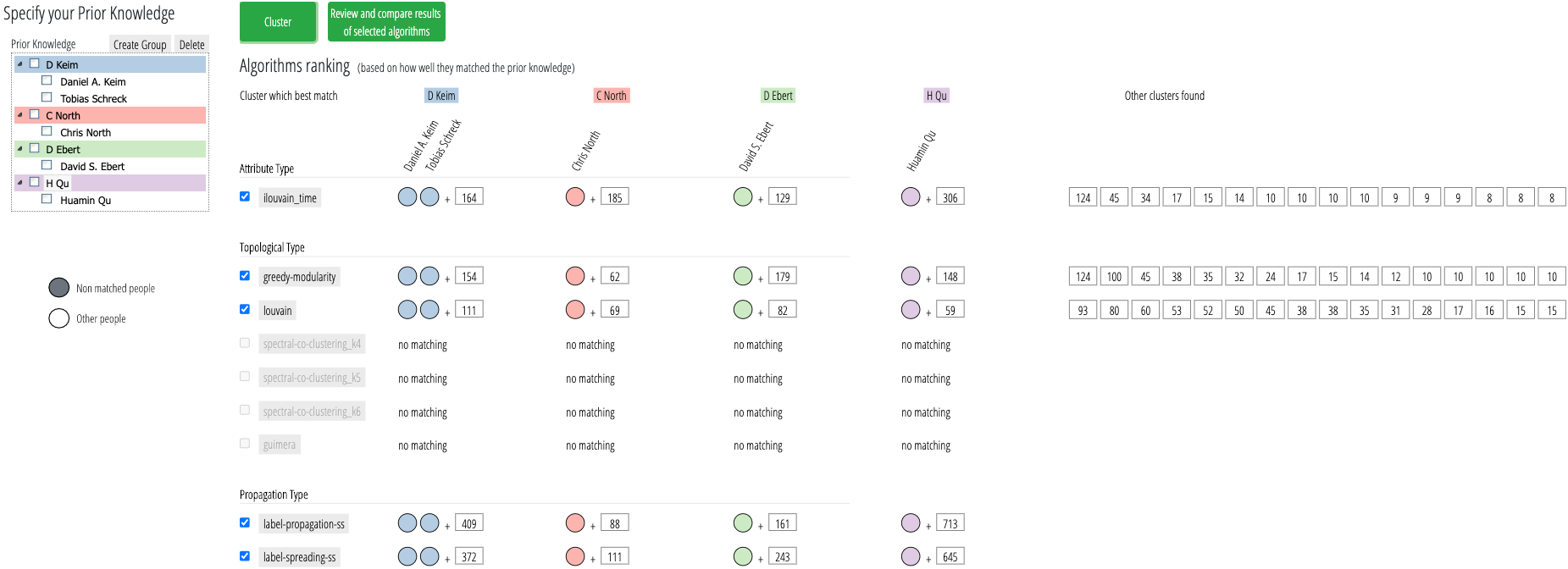}    
    \caption{Computing the Lineages of VAST authors: Prior Knowledge from Alice and results of the clusterings matching it.}
    \label{fig:Vast_PK_Clustering_View}
\end{figure}

In the second cases study we took the role of Alice, a VAST Steering Committee (SC) member, who participates in a SC meeting to validate the Program Committee proposed by the VAST paper chairs for the next conference.
One of the many problems that all conference organizers face is to balance the members of the Program Committee according to several criteria.
The InfoVis Steering Committee Policies FAQ states that the composition of the Program Committee should consider explicitly how to achieve an appropriate and diverse mix ~\cite{infovisfaq} of:    
\begin{itemize*}
\item academic lineages
\item research topics
\item job (academia, industry)
\item geography (in rough proportion to the research activity in major regions)
\item gender.
\end{itemize*}
Most of these criteria are well understood, except \emph{academic lineage} which is not clearly defined.
Alice will use the ``Visualization Publications Data'' (VisPubData~\cite{VisPubData}) to find-out if she can objectify this concept of lineage to check the diversity of the proposed Program Committee accordingly.

Using PK-clustering, Alice loads the VisPubData, filtered to only contain articles from the VAST conference, between 2009--2018. Only prolific authors can be members of the program committee, but highly filtering the co-authorship network  would change its structure and disconnect it. Thus, she will use the unfiltered network of 1383 authors to run the algorithms and perform the matching (Step 1 of the process), even if at the end only 113 authors with more than 4 articles will be need to consolidated (Steps 2 and 3).

Alice starts the PK-clustering process by entering her prior knowledge, which is partial and based on two strategies: her knowledge of some areas of VAST, and the name of well-known researchers who have developed their own lineage. She runs the algorithms (\autoref{fig:Vast_PK_Clustering_View}) and 5 algorithms produce a perfect match, acknowledging her knowledge of some areas of VAST. She then shows the results to other members of the SC who will help her consolidate the lineage clusters.

\begin{figure}[tb]
    \centering
    \includegraphics[width=\linewidth]{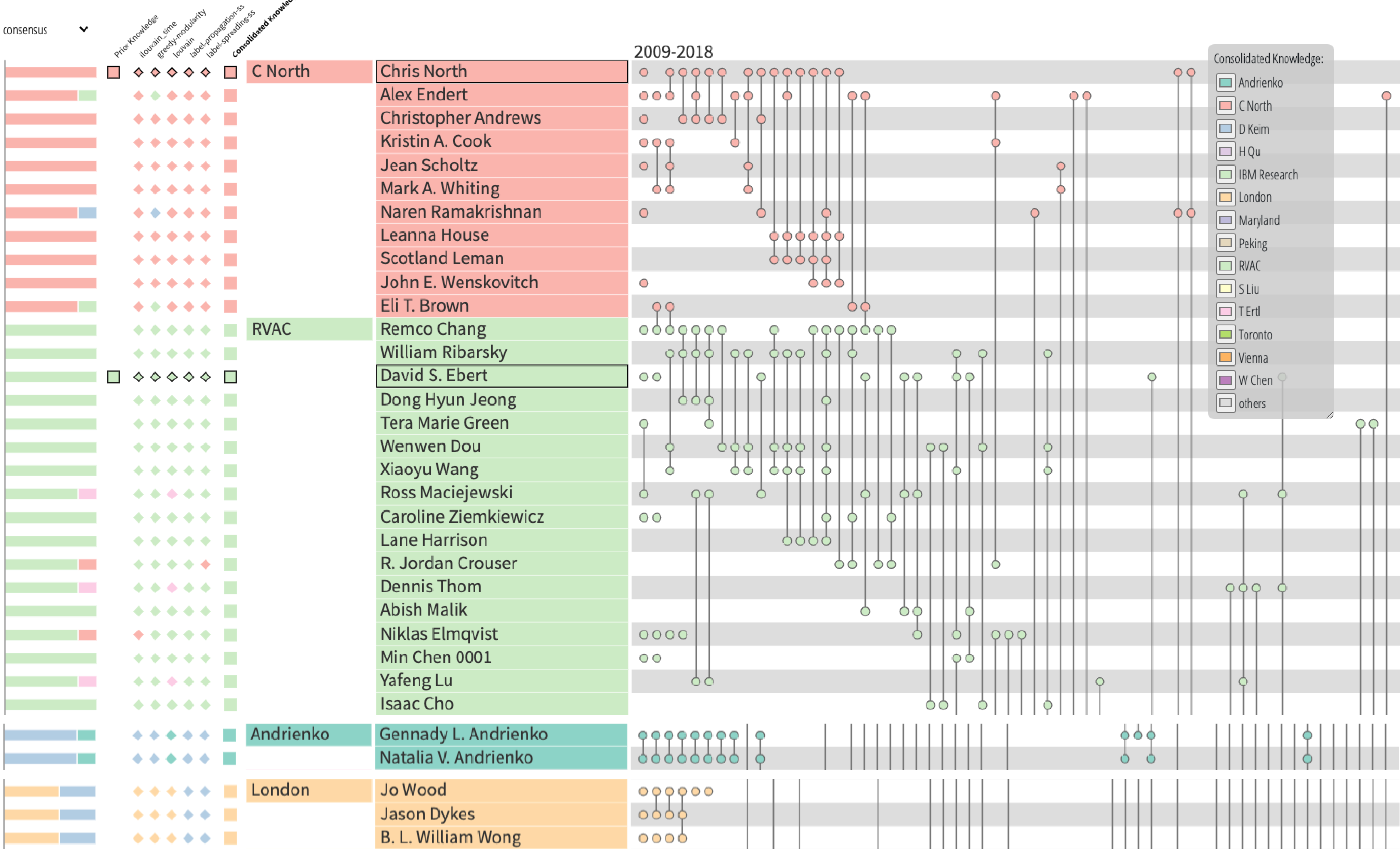}    
    \caption{Four consolidated groups in the VAST dataset: C North, RVAC, Andrienko and London}
    \label{fig:vast_some_consolidated_groups}
\end{figure}

Her initial PK clusters are quickly consolidated, using Internet search to validate some less known authors. She then decides to create as many additional clusters and lineage groups as she can. 
For some authors, she decides to override the consensus of the algorithms. For example, she decides, and her colleagues agree, that Gennady and Natalia Andrienko should be in their own lineage group and not in D. Keim's (\autoref{fig:vast_some_consolidated_groups}). The history of VAST in Europe, very much centered around D.\ Keim and the VisMaster project~\cite{VisMaster}, has strongly influenced the network structure and some external knowledge is required to untangle it.

Using the \emph{PK\_louvain} algorithm as starting point, Alice creates new groups and achieves a consensus among the experts on a plausible set of lineages for VAST.
She then checks with the list proposed by the program committee by entering it in on a spreadsheet with the names and affiliations. She adds the groups and their color, and sort the list by group.
Alice can now report her work to the whole SC, which can check the balance of lineages according to this analysis, and decide if some lineage groups are over or under represented.  By keeping the affiliations in the list, the SC can also check the balance of affiliations that is not always aligned with the lineages.  The final results are available in the supplemental material of the article.


Using partitioning clustering (although with outliers) forces the algorithms or experts to make strong decisions related to lineages. But using a soft clustering (or overlapping partitions), while providing a more nuanced view of lineages, would not be as simple to interpret as coloring spreadsheet lines and sorting them;
in the end, the final selection only uses the lineage criterion among many others.  Still, we believe PK-clustering can provide a partial but concrete answer to the problem of defining what the scientific lineages are.

\subsection{Feedback from practitioners}

Although we could not conduct face to face meetings with historians and sociologists due to the COVID19 lockdown, we showed the system to three practitioners and asked their feedback through videoconferencing systems, sharing video demonstrations and sharing our screen. 

They all acknowledged the pitfalls of existing systems providing clustering algorithms as black boxes with strange names and mysterious parameters. 
They also agreed that the current process for clustering a social network was cumbersome when they wanted to validate the groups and compare the results of different algorithms. None of the popular and usable systems provide easy ways to compare the results of the clusterings. Usually, the analyst needs to try a few algorithms, remembering the groups that seemed good in some of the algorithms, sometimes printing the clustered networks to keep track of the different options. Still, they all confirmed that they usually stop after trying 2 to 3 algorithms because of lack of time and support from the tools. Evaluation of clusterings is long and tedious.

They were intrigued by the idea of entering the prior knowledge to the system, but acknowledged that it was easy to understand and natural for them to think in terms of well-known entities belonging to groups. They felt uneasy thinking that this prior knowledge could bias the results of the clustering and of the analysis. However, after a short discussion, they also agreed that the traditional process of picking in a more or less informed way two or three algorithms to perform a clustering was also probably priming them and adding other biases. Still, they said that they would need to explain the process clearly in their publications and that some reviewers could also stress the risks.

They all agreed that the process was clear and made sense, but they also felt it was complicated and that they would need time to master it. They said that it was more complicated than pressing a button, but that the extra work was worth it.

One historian who spends a lot of time analyzing her social networks and finding information about all the people was shocked by the idea that you could want to use an algorithm that did not match fully the prior knowledge. For us, it matters if the prior knowledge is given as constraints or preferences, but we did not want to introduce these notions in the user interface so analysts are free to interpret the prior knowledge as one or the other.

They also identified some issues with the prototype. It was not managing disconnected networks at all when we showed the demo, and they stressed the fact that real networks always have disconnected components. They were also asking about structural transformations, such as filtering by attribute or by node type. We chose not support these functions at this stage, but they can be done through other standard network systems.

They were also interested in getting explanations about the algorithms, why some would pick the right groups and others would not. Our system is not meant to provide explanations and works with black box algorithms. We wished we could help them but that would be another project. Still, when an attribute-based algorithm matches the prior knowledge, we believe that attribute-based explanations are more understandable, \eg groups based on time, or income.

\rev{The table and summary report was added after those sessions so no feedback was gathered}.  We will continue to collaborate with those practitioners and help them test PK-culstering during their next social network analysis project.


\section{Discussion}

\begin{revs}
As presented in~\autoref{sec:summary}, the existing approaches to create clusters in social networks consider three options: standard clustering, ensemble clustering, and semi-supervised clustering. Our proposed PK-clustering approach combines aspects of the three options in order to give more control to users in the analysis loop, and allow them to have more say in the final results. 

Proponents of automatic methods may argue that PK-clustering gives users too much influence on the final result as they can change the cluster assignments at will. On the other hand we know that social scientists are rarely satisfied with current clustering methods, in part because they run on graph data that rarely represent all the knowledge they have of the social network, so providing user control to correct mistakes is critical.

Traditional methods push users to believe the results of the first algorithms and parameter selection they try (typically chosen randomly). Using PK-Clustering, users can still follow blindly the results of one algorithm but PK-clustering provides a more systematic approach. It allows users to compare results, review consensus, think at each phase and reflect on decisions. Instead of passively accepting what the algorithms propose, users provide initial hypotheses---which limits the chances of being primed by an algorithm, and explicitly validate the cluster assignment of nodes, therefore performing a critical review of the automated results, yet with fast interaction to accept many suggestions at once when appropriate.

This new approach allows users to discover alternative views. For example when algorithms do not match the PK, it is an indication that the PK is being challenged and may not be correct.  Users actively participate in the process of assigning, a requirement for social scientists. The report produced at the end of the analysis adds transparency by recording where the results come from for each node so decisions can be reviewed. Ultimately social scientists remain responsible for reporting and justifying their choices and interventions in their publication.

We acknowledge that bias issues are complex. The absence of ground truth limits researchers' ability to measure those biases, and no approach solves all issues yet, but we believe that PK-clustering offers a fresh perspective on those issues and will lead to results that are more useful to social scientists.
\end{revs}

\subsection{Limitations}

Many more clustering algorithms exist and could be added. 
Moreover, expanding the exploration of parameter spaces for clustering algorithms seems needed.  Another limitation of the current prototype is that some algorithms do not work well with disconnected components of the graph. Unfortunately, social scientists datasets typically have many disconnected components. This issue can be mitigated by separating components into a set of connected components, run the algorithms on them, and merge the results.
Our prototype runs both with node-link and PAOH representations, but it is better tuned to the PAOH representation because of its highly readable nodes list and table format which makes the review of consensus easier. Better coordination of the table with node link diagrams and other network visualizations is needed. 
\rev{Further case studies will help us improve the utility of the tool as well as the provenance table and summary, which could include annotations documenting the decision process}

\subsection{Performance}
\label{sub:performances}

\rev{The performance of PK-clustering strongly depends on the clustering algorithms. We implemented fast algorithms to have acceptable computation times.  Currently a cut-off automatically removes algorithms that
have not produced a clustering after 10 seconds of computation.
We ran a benchmark of the performance on the two datasets of the case studies with a laptop equipped with an Intel Core i7-8550U CPU 1.80GHz × 8 and 16 Gigabytes of memory.
For the full Marie Boucher social network described in \autoref{sud:MB}, composed of 189 nodes and 58 hyperedges (1000 edges after the unipartite projection) it took 0.6 seconds to run all our implemented algorithms and produce the matching.
For the graph of \autoref{sud:VAST} about the VisPubData of the VAST conference, made of 1383 nodes and 512 hyperedges (4554 edges after projection), one algorithms (the Label Propagation algorithm) took 11.37 seconds to finish and was  abandoned because deemed too computationally expensive. 
Those two datasets are representative of the many medium size datasets historians and social scientists carefully curate (\ie 50--500 nodes).}


In order to improve the computational scalability, we will implement progressive techniques to deal with larger sizes~\cite{Progressive}. 
The current user interface design for PK-clustering would allow the ranked list of algorithms to be progressively updated, and users to review a few individual algorithms first while other algorithms are still running.  Of course, visual scalability is also an issue with larger datasets, as the list of people also grows. 
PAOHVis allows groups (like clusters) to be aggregated or expanded, so we expect that users would expand clusters one by one to review and consolidate them, while also being able to review the connections between the proposed clusters.  \rev{Users can also use the automated features of PK-Clustering to consolidate the nodes (\eg selecting one algorithm based on the ranking, or using the consensus slider to consolidate all the nodes at once). Pixel-oriented visualizations~\cite{keim2000pixel} would facilitate the review of consensus for a large number nodes and clusters.}
Classic techniques like zooming or fisheye views~\cite{Jakobsen06-fisheye, rao94} would help as long as names remain readable, which is critical to our users.

\section{Conclusion}

In this article, we introduced a new \rev{approach}, called PK-clustering, to help social scientists create meaningful clusters in social networks. It is composed of three phases: 1) users specify the prior knowledge by associating a subset of nodes to groups, 2) all algorithms are run and ranked, 3)  users review and compare results to consolidate the final clusters.

This mixed-initiative approach is more complex than a traditional clustering process where users simply press a button and get
the results, but it provides social scientists with an opportunity to correct mistakes and infuse their deep knowledge of the people and their lifes in the results. With simple actions such as moving a slider, or dragging over icons, users are able to interactively perform complex tasks on many nodes at once. The output of PK-clustering is---using a direct quote from a social scientist providing feedback on the prototype:  ``a clustering that is supported by algorithms and validated, fully or partially, by social scientists according to their prior knowledge''. Two case studies illustrated the benefits of PK-clustering.

\rev{Clustering and social network analysis remains a challenging task, typically without ground truth to formally evaluate the results. The risk of introducing bias remains always present, in this new approach as well as in traditional methods. We believe that PK-clustering offers a fresh perspective on the process of clustering social networks and gives users the opportunity to report their results in a transparent manner. }The next frontier will be the analysis of dynamic social networks, that are often used in social science, and our approach will need to take into account the evolution of the communities over time.

\acknowledgments{
The authors wish to thank Nicole Dufournaud, Pascal Cristofoli, Francesco Napolitano, and Christophe Prieur. This work was supported in part by a grant from he DataIA Institute (project HistorIA) and the European project IVAN.}

\bibliographystyle{abbrv-doi-hyperref}
\balance
\bibliography{main}
\end{document}